\documentstyle[12pt]{amsart}

\catcode`\@=11

\long\def\@savemarbox#1#2{\global\setbox#1\vtop{\hsize\marginparwidth 
  \@parboxrestore\tiny\raggedright #2}}
\newcommand\lref[1]{\ref{#1}%
\@ifundefined{r@DisplaY #1}{}{ (#1)}}

\newcommand\fakelabel[2]{\@bsphack\if@filesw {\let\thepage\relax
   \newcommand\protect{\noexpand\noexpand\noexpand}%
\xdef\@gtempa{\write\@auxout{\string
      \newlabel{#1}{{#2}{\thepage}}}}}\@gtempa
   \if@nobreak \ifvmode\nobreak\fi\fi\fi\@esphack}

\catcode`\@=12

\def\Empty{}
\newcommand\oplabel[1]{
  \def\OpArg{#1} \ifx \OpArg\Empty {} \else
        \label{#1}
  \fi}
\newtheorem{theoremSt}{Theorem}[section]

\newtheorem{exampleSt}[theoremSt]{Example}
\newtheorem{exerciseSt}[theoremSt]{Exercise}

\newcommand\MakeStEnv[1]{
  \newenvironment{#1}[1]{
  \begin{#1St} \oplabel{##1}%
  \global\def\CrntSt{\thetheoremSt}%
}{ 
  \end{#1St} }
  \newenvironment{#1+}[1]{
  \begin{#1St} \label{##1}%
  \label{DisplaY ##1}%
  \global\def\CrntSt{\thetheoremSt}%
  \def\Labl{##1}\ifx\Labl\Empty{} \else {\em (\Labl)\,}\fi%
}{ 
  \end{#1St} }
}
\MakeStEnv{theorem}
\MakeStEnv{corollary}
\MakeStEnv{proposition}
\MakeStEnv{lemma}
\MakeStEnv{definition}
\MakeStEnv{conjecture}

\newlength{\saveu}

\newcommand{\startproof}[1]{%
\medbreak\mbox{}\noindent{\it Proof of #1:}%
}
\newcommand{\finishproof}[1]{ 
  \def\FPArg{#1}
  \ifx\FPArg\Empty
        \newcommand\FPArg{\CrntSt}  \fi
  \smallbreak\noindent\makebox[\textwidth]{\hfill\fbox{\FPArg}}
  \medbreak\noindent
}

\newcommand\FF{{\cal F}}

\newcommand\LL{{\cal L}}
\newcommand\MM{{\cal M}}

\newcommand\PP{{\cal P}}

\newcommand\PMF{{\PP\kern-2pt\MM\FF}}

\newcommand\PML{{\PP\kern-2pt\MM\LL}}

\newcommand\bbR{{\mathord{\text{I\kern-2pt R}}}}        
\newcommand\bbH{{\mathord{\text{I\kern-2pt H}}}}        

\newcommand\bigrightarrow[1]{\hbox to #1{\rightarrowfill}}
\newcommand\bigleftarrow[1]{\hbox to #1{\leftarrowfill}}

\newcommand\semidir{\mathrel{\hbox{\vrule depth-.03ex height1.1ex\kern-0.15em$\times$}}}

\numberwithin{equation}{section}

\begin{document}

\title[Mass and Decay Spectra of the Piecewise Uniform String]
{Mass and Decay Spectra of the Piecewise Uniform String}  
\author{I. Brevik}
\address{Applied Mechanics, Norwegian University of Science and Technology,
N-7034 Trondheim, Norway\,\,{\em E-mail address:}
{\rm Iver.H.Brevik@mtf.ntnu.no}}
\author{A.A. Bytsenko}
\address{Departamento de Fisica, Universidade Estadual de Londrina,
 Caixa Postal 6001, Londrina-Parana, Brazil\,\, {\em E-mail address:} 
 {\rm abyts@fisica.uel.br}}
\author{A.E. Gon\c calves}
\address{Departamento de Fisica, Universidade Estadual de Londrina,
 Caixa Postal 6001, Londrina-Parana, Brazil\,\, {\em E-mail address:} 
 {\rm goncalve@fisica.uel.br}}

\date{February, 1999}


\thanks{Second author partially supported by a CNPq grant (Brazil), RFFI 
grant (Russia) No 98-02-18380-a, and by GRACENAS grant (Russia) No 6-18-1997.}

\maketitle

\begin{abstract}

Mass and decay spectra are calculated for quantum massive
exitations of a piecewise uniform bosonic string. The physical meaning of the
critical temperatures characterising the radiation in the decay of a massive
microstate in string theory is discussed.
\end{abstract}

\vspace{1.0cm}

The composite string, in which the (relativistic) string is
assumed to consist of two (or more) separately uniform pieces, is a variant
of the conventional theory. This theory has been generalized
and further studied from various points of view \cite{brevik90} -
\cite{brevik98h}. The composite
string model may serve as a useful two-dimensional field theoretical model.
Usually, a two-dimensional field theory describes a 
particular classical solution of string theory by constructing a matter
system. The vanishing total central charge of a system ensures the
existence of a BRST operator, playing a crucial role in world-sheet
and space-time gauge invariance. Such an operator can be interpreted
as a generator of linearized gauge transformations, mixing ghosts and matter.

It is very important that two-dimensional topological field theories (like
ordinary string models) can sometimes be given space-time interpretations
for which the usual decoupling of ghosts and matter does not hold. For example,
three-dimensional Chern-Simons gauge theory can arise as a string theory 
\cite{witten}.
Relations between gauge fields and strings present fascinating and unanswered
questions. The full answer to these questions is of great importance for
theoretical physics. It will provide the true gauge degrees of freedom
of the fundamental string theories, and therefore also of gravity 
\cite{gubser}.

{\bf 1.} In this paper we consider the motion
of a two-piece classical string in flat $D$-dimensional space-time.
Following the notation in \cite{green87}
we let $X^\mu(\sigma,\tau)\,\,\,(\mu = 0,1,2,\cdots
(D-1))$ specify the coordinates on the world sheet. 
The action has the form

$$
S = -\frac{1}{2}\int d\tau d\sigma T(\sigma) \eta^{\alpha \beta} 
\partial_\alpha X^\mu
\partial_\beta X_\mu
\mbox{,}
\eqno{(1)}
$$
where $T(\sigma)$, the position-dependent tension
$T(\sigma) = T_{\rm I} + (T_{\rm II} - T_{\rm I}) \theta(\sigma - L_{\rm I})$, 
contains
the step function $\theta(y > 0) = 1,\,\,\,\theta(y < 0) = 0$.
The tensions $T_{\rm I}$  and  $T_{\rm II}$ are associated with the pieces 
$L_{\rm I}$ and $L_{\rm II}$ of the string. 
The momentum conjugate to $X^\mu$ is $P^\mu(\sigma) = T(\sigma) \dot{X}^{\mu}$.
The Hamiltonian of the two-dimensional sheet becomes accordingly

$$
H = \int_{0}^{\pi} \left[ P_\mu(\sigma) \dot{X}^\mu -{\frak L} 
\right] d\sigma =
\frac{1}{2} \int_{0}^{\pi} T(\sigma) (\dot{X}^2 + {X'}^2)d\sigma
\mbox{,}
\eqno{(2)}
$$
where ${\frak L}$ is the Lagrangian.
The basic condition that we shall impose, is that $H = 0$ when
applied to the physical states. This is a more weak condition than
the strong condition $T_{\alpha \beta} = 0\,\,\,(\alpha,\beta=0,1)$ 
on the energy-momentum tensor, applicable for a uniform string.

{\bf 2.} We quantize the system according to conventional methods as
presented in Ref. \cite{brevik98} (see for detail Ref. \cite{green87}). 
In accordance with the canonical prescription in region I the 
equal-time commutation rules are required to be

$$
T_{\rm I}[\dot{X}^\mu(\sigma,\tau), X^\nu(\sigma', \tau)]=
-i \delta(\sigma-\sigma')\eta^{\mu \nu}
\mbox{,}
\eqno{(3)}
$$
and in region II

$$
T_{\rm II} [\dot{X}^\mu(\sigma,\tau),X^\nu(\sigma',\tau)] =
-i \delta(\sigma - \sigma')\eta^{\mu \nu}
\mbox{,}
\eqno{(4)}
$$
where $\eta^{\mu \nu}$ is the $D$-dimensional metric. These relations are in
conformity with the fact that the momentum conjugate to $X^\mu$ is in either
region equal to $T(\sigma) \dot{X}^\mu$. The remaining commutation relations
vanish:

$$
[X^\mu(\sigma,\tau),X^\nu(\sigma',\tau)] = [\dot{X}^\mu(\sigma,\tau),
\dot{X}^\nu(\sigma',\tau)] = 0
\mbox{.}
\eqno{(5)}
$$

The quantities to be promoted to Fock state operators are
$\alpha_{\mp n}(s)$ and $\tilde{\alpha}_{\mp n}(s)$ (first branch, region I),
$\gamma_{\mp n}(s)$ (first branch, region II),
$\alpha_{\mp n}(s^{-1})$ and $\tilde{\alpha}_{\mp n}(s^{-1})$ (second branch,
region I), and $\gamma_{\mp n}(s^{-1})$ (second branch, region II), where
$s=L_{\rm II}/L_{\rm I}$. These operators satisfy

$$
\alpha_{-n}^{\mu}(s) = \alpha_{n}^{\mu\dagger}(s),\,\,\,
\gamma_{-n}^{\mu}(s) = \gamma_{n}^{\mu\dagger}(s)
\mbox{,}
$$
$$
\alpha_{-n}^{\mu}(s^{-1}) = \alpha_{n}^{\mu\dagger}(s^{-1}),\,\,\,
\gamma_{-n}^{\mu}(s^{-1}) = \gamma_{n}^{\mu\dagger}(s^{-1})
\mbox{,}
\eqno{(6)}
$$
for all $n$. 

For the first branch we then get 

$$
[\alpha_n^\mu(s),\alpha_m^\nu(s)] = n \delta_{n+m,0} \eta^{\mu\nu},\,\,\,\,\,
[\gamma_{n}^\mu(s),\gamma_{m}^\nu(s)] = 4nx \delta_{n+m,0} \eta^{\mu\nu}
\mbox{,}
\eqno{(7)}
$$
with a similar relation for the $\tilde{\alpha}_n$. Here we put 
$x=T_{\rm I}/T_{\rm II}$.
For the second branch we get analogously

$$
[\alpha_{n}^{\mu}(s^{-1}),\alpha_{m}^{\nu}(s^{-1})]=n\delta_{n+m,0}
\eta^{\mu\nu},
\,\,\,\,\,
[\gamma_{n}^{\mu}(s^{-1}), \gamma_{m}^{\nu}(s^{-1})]= 4nx\delta_{n+m,0}
\eta^{\mu\nu}
\mbox{.}
\eqno{(8)}
$$

By introducing annihilation and creation operators for the first branch in the
following way:

$$
\alpha_n^\mu(s) =  \sqrt{n} a_n^\mu(s),\,\,\,\,\,
\alpha_{-n}^\mu(s)=\sqrt{n} a_n^{\mu\dagger}(s)
\mbox{,}
$$
$$
\gamma_n^\mu(s) = \sqrt{4nx} c_{n}^{\mu}(s),\,\,\,\,\,
\gamma_{-n}^\mu(s) =
\sqrt{4nx} c_{n}^{\mu\dagger}(s)
\mbox{,}
\eqno{(9)}
$$
we find for $n \geq 1$ the standard form

$$
[a_{n}^{\mu}(s), a_{m}^{\nu\dagger}(s)] = 
[c_{n}^{\mu}(s), c_{m}^{\nu\dagger}(s)]=\delta_{nm}\eta^{\mu\nu}
\mbox{.}
\eqno{(10)}
$$
The commutation relations for the second branch are analogous,
only with the replacement $s\rightarrow s^{-1}$.
In the following we shall limit ourselves to the first branch only.
The Hamiltonians and mass are given by 

$$
H_{\rm I} = -\frac{M^2x}{2st(s)} + \frac{1}{2}\sum_{n=1}^{\infty} \omega_n(s)
[a_n^{\dagger}(s)\cdot a_n(s) + \tilde{a_n}^{\dagger}(s)\cdot\tilde{a_n}(s)]
\mbox{,}
\eqno{(11)}
$$
$$
H_{\rm II} = -\frac{M^2}{2t(s)} + s \sum_{n=1}^{\infty} \omega_n(s)
c_n^{\dagger}(s) \cdot c_n(s)
\mbox{,}
\eqno{(12)}
$$

$$
M^{2} = t(s)\sum_{i=1}^{24}\sum_{n=1}^{\infty}\omega_n(s)
[a_{ni}^{\dagger}(s)a_{ni}(s) + \tilde{a}_{ni}^{\dagger}\tilde{a}_{ni}(s)
- 2] 
$$
$$
+ 2st(s)\sum_{i=1}^{24}\sum_{n=1}^{\infty}\omega_n(s)
[c_{ni}^{\dagger}(s)c_{ni}(s)- 1]
\mbox{.}
\eqno{(13)}
$$
where $\omega_n(s)=(1+s)n$, $t(s)=\pi T_{\rm II}s/(s+1)$ and we have used the 
notation $c_n^{\dagger}c_n\equiv c_n^{\mu\dagger}c_{n\mu}$.
We have here put $D = 26$, the commonly accepted space-time dimension for the
bosonic string.
As usual, the $c_{ni}$ denote the transverse oscillator operators  (here for
the first branch).

The free energy
of the field content in the "proper time" representation can be 
written as follows \cite{brevik98}

$$
F=-\frac{1}{24}(s+\frac{1}{s}-2)-2^{-40}\pi^{-26}t(s)^{-13}
$$
$$
\times\int_0^{\infty}\frac{d\tau_2}{\tau_2^{14}}
\int_{-1/2}^{1/2}d\tau_1\left[\theta_3\left(0|\frac{i\beta^2t(s)}
{8\pi^2\tau_2}\right)-1\right]
$$
$$
\times|\eta[(1+s)\tau]|^{-48}\eta[s(1+s)(\tau-\overline{\tau})]^{-24}
\mbox{,}
\eqno{(14)}
$$
where we integrate over all possible non-diffeomorphic toruses which are
characterized by a single Teichm{\"u}ller parameter $\tau=\tau_1+i\tau_2$.
In Eq. (14) the Dedekind $\eta$-function and the Jacobi $\theta_3$-function

$$
\eta(\tau) = e^{\frac{\pi i\tau}{12}}\prod_{n=1}^{\infty}(1 - e^{2\pi in\tau})
\mbox{,}
\eqno{(15)}
$$
$$
\theta_3(v|x) = \sum_{n=-\infty}^{\infty} e^{ixn^2+2 \pi ivn}
\mbox{,}
\eqno{(16)}
$$
and the condition $\eta(-\overline{\tau})=\overline{\eta(\tau)}$ has been
used. Once the free energy has been found, the other thermodynamic quantities 
can readily be calculated. For instance, the energy $U$ and the entropy $S$ of
the system are $U = \partial (\beta F)/\partial \beta,\,\,
S =  \beta^2 \partial F/\partial \beta$.

The integrand in Eq. (14) is ultraviolet finite if

$$
\beta > \beta_c = \frac{4}{s}\sqrt{\frac{\pi(1+s)}{T_{\rm II}}}
\mbox{.}
\eqno{(17)}
$$
For a fixed value of $T_{\rm II}$ the Hagedorn temperature is thus depends
on $s$. 
Finally, let us consider the limiting case in which one of the pieces of the
string is much shorter than the other. Physically this case is of interest,
since it corresponds to a point mass sitting on a string. Since we have
assumed that $s \geq 1$, this case corresponds to $s \rightarrow\infty$. We
let the tension $T_{\rm II}$ be fixed, though arbitrary.
It is seen, first of all, that the Hagedorn temperature (17) goes to
infinity so that $F$ is always ultraviolet finite,
$\beta_c \rightarrow 0,\,\,T_c \rightarrow \infty$.
Next, since $\exp\left(-\beta^2t(s)/8\pi^2\tau_2\right)$ can be
taken to be small we obtain

$$
F_{(\beta\rightarrow 0)} = -\frac{s}{24}-
(8\pi^3T_{\rm II})^{-13}\int_0^{\infty}\frac{d\tau_2}{\tau_2^{14}}
\int_{-1/2}^{1/2}d\tau_1
$$
$$
\times  \exp\left(-\frac{\beta^2T_{\rm II}}{8\pi\tau_2}\right)
|\eta[(1+s)\tau]|^{-48}\eta[s(1+s)(\tau-\overline{\tau})]^{-24}
\mbox{.}
\eqno{(18)}
$$
Physically speaking, the linear dependence of the first term in Eq. (18)
reflects that the Casimir energy of a little piece of string
embedded in an essentially infinite string has for dimensional reasons to be
inversely proportional to the length $L_I = \pi/(1+s) \simeq \pi/s$ of
the little string. The first term in (18) is seen to outweigh the
second, integral term, which goes to zero when $s \rightarrow \infty$.

{\bf 3.} The spectrum for the decay of a massive initial state 
$|\Phi>$ of momentum
$p_{\mu}=(M,{\bf 0})$ into state $|\Phi_X>\otimes|k_{\mu}>$ with momentum 
$k_{\mu}\,\,(k^2=0)$ can be
presented by the modulus squared of the amplitude summed over all state
$|\Phi_X>$.
The initial state $|\Phi>$ in string theory characterized by a partition 
$\{N_n\}$ of $N=\sum_{n=1}^{\infty}\omega_nN_n$, where 
$N_n=\sum_{j}\alpha_{nj}^{\dagger}\alpha_{nj}$ is the occupation number 
of the $n-$ th mode.
The inclusive photon spectrum for the decay of state $\Phi_{\{N_n\}}$
is given by the sum over all states $\Phi_{X}$ satisfying the mass condition
and has the form \cite{amati99}

$$
d\Gamma_{\Phi\{N_n\}}(k_0)=\sum_{\{\Phi_X\}}|<\Phi_X|V(k)|\Phi_{\{N_n\}}>|^2
{\rm Vol}({\Bbb S}^{D-2})k_0^{D-3}dk_0
\mbox{.}
\eqno{(19)}
$$
Here ${\rm Vol}({\Bbb S}^{D-2})=2\pi^{(D-1)/2}[\Gamma((D-1)/2)]^{-1}$,
the photon vertex operator $V(k)$ is given by
$V(k)=\xi_{\mu}\partial_{\tau}X^{\mu}(0)\exp(ik.X(0))$,
$\xi_{\mu}$ is the photon polarization vector and $\xi_{\mu}k^{\mu}=0$.
It can be shown that non-planar contribution does not change the above result
\cite{amati99}.

Let $m_0$ and $\omega_0$ be the Kaluza-Klein momentum and winding number 
associated with the state of closed string compactified on a circle of radius
$R$. For the piecewise uniform string the calculation of spectrum factorizes 
into left-right parts of branch I, and a branch II (note also that
$[H_{\rm I}, H_{\rm II}]=0$). Following the calculation presented in
\cite{amati99} we obtain the final result:

$$
d\Gamma_{\Phi({\rm I,II})}(k_0)={\cal C}\xi^2M^2\prod_{\ell=1}^3
\left[\frac{e^{-\frac{k_0}{T_\ell}}}{1-e^{-\frac{k_0}{T_\ell}}}\right]
{\rm Vol}({\Bbb S}^{D-2})k_0^{D-1}dk_0
\mbox{,}
\eqno{(20)}
$$
where ${\cal C}$ is some constant. For the region II the critical temperature
is $T_1=T_c=\beta_c^{-1}$, while for the region I the temperatures related
with right- and left-moving modes are correspondingly given by
$T_2=2(M^2-Q_{+}^2)^{1/2}/(aM),\,\, T_3=2(M^2-Q_{-}^2)^{1/2}/(aM)$, where
$Q_{\pm}=m_0/R\pm \omega_0R$ and $a=2\pi[(D-2)/6]^{1/2}$.

The decay spectrum can be given by analogy with black holes \cite{maldacena}:

$$
d\Gamma_{\Phi({\rm I,II})}(k_0)={\cal C}\xi^2M^2\Sigma(k_0)
\frac{e^{-k_0/T}}{1-e^{-k_0/T}}
{\rm Vol}({\Bbb S}^{D-2})k_0^{D-2}dk_0
\mbox{,}
\eqno{(21)}
$$
$$
\Sigma(k_0)=k_0e^{-k_0/T_1}\left(1-e^{-k_0/T}\right)\prod_{\ell=1}^3
\left[1-e^{-k_0/T_{\ell}}\right]^{-1}
\mbox{,}
\eqno{(22)}
$$
the factor $\Sigma(k_0)$ depends on temperature $T$, where 
$T^{-1}=T_2^{-1}+T_3^{-1}$. 
The radiative spectrum from microscopic string states related with
region I and II is exactly thermal.
When $M\simeq Q_{+}$, $T_2\simeq T_3=2\left(M^2-Q_{+}^2\right)^{1/2}/(aM)$
radiation vanishes. The same situation occur when $s\simeq 0\,\,(T_c\simeq 0)$.

Our fundamental theory is based upon the following two conditions at the 
junctions: the transverse displacement
itself, as well as the transverse force, has to be continuous. Moreover, 
we make two simplifying assumptions: first,
the tension ratio $x$ is taken to be small, 
$x \rightarrow 0$. This assumption implies that the eigenvalue spectrum 
for the composite string becomes quite simple: there are two branches, 
the first branch corresponding to 
$\omega_n(s)=(1+s)n$ while
the second branch corresponding to 
$\omega_n(s^{-1})=(1+s^{-1})n$,
with $n$ an integer. Our second assumption is that $s$ is an integer.

We consider the first branch only.  The right- and left-moving amplitudes in 
region I can be chosen freely, while the amplitudes in region II are 
thereafter fixed. The oscillations in region I are therefore as for 
a {\it closed} string, whereas the oscillations in region II are standing 
waves, corresponding to an {\it open} string. This is the physical background 
why there is only one single critical temperature $T_c$ in region II in 
Eqs.(20) and (22), while there are two critical temperatures $T_2$ and $T_3$ in 
region I.
It should be emphasized once more that our composite string is 
{\it relativistic}, in the sense that the transverse velocity of sound is 
everywhere in the string equal to velocity of light. How to construct a 
composite string 
theory in the absence of this relativistic requirement is not known.

Finally we note that the black hole entropy behaviour can be understood in 
terms of the degeneracy of some interacting fundamental string excitation
mode. Generally speaking the fundamental string and $p$-brane approach
can yield  a microscopic interpretation of the entropy. It will be 
interesting to consider phases of the piecewise uniform (super) string and
the Bekenstein- Hawking entropies of black holes associated with this
string. The entropies can be derived by counting black hole microstates;
the laws of black hole dynamics could then be identified with the laws of
thermodynamics. We hope that the proposed calculation will be of interest 
in view of future applications to concrete problems in string theory,
quantum gravity and in black hole physics.

\end{document}